\documentclass[lettersize,journal]{IEEEtran}
\usepackage{amsmath,amsfonts}
\usepackage{algorithmic}
\usepackage{algorithm}
\usepackage{array}
\usepackage[caption=false,font=normalsize,labelfont=sf,textfont=sf]{subfig}
\usepackage{textcomp}
\usepackage{stfloats}
\usepackage{url}
\usepackage{verbatim}
\usepackage{graphicx}
\usepackage{cite}
\usepackage{adjustbox, diagbox, caption, booktabs}
\usepackage{tabularx, multirow, multicol, makecell}
\usepackage{hyperref}
\usepackage{mdframed}
\usepackage{xcolor}
\definecolor{redbrown}{rgb}{0.65, 0.16, 0.16}

\newcommand{\squad}{\hspace{0.3em}} 

\newcommand{\floor}[1]{\lfloor #1 \rfloor}

\definecolor{lightblue}{rgb}{0.93, 0.96, 0.98}

\begin{document}

\title{Listen, Chat, and Remix: Text-Guided Soundscape Remixing for Enhanced Auditory Experience}

\author{Xilin Jiang, Cong Han, Yinghao Aaron Li, Nima Mesgarani
\thanks{All with Department
of Electrical Engineering, Columbia University, New York,
NY, USA. Email: xj2289@columbia.edu; nima@ee.columbia.edu. \\This work involved human evaluation. Approval of all
ethical and experimental procedures and protocols was granted by the Columbia
University’s Institutional Review Board (IRB protocol number AAAR8655).}
}

\markboth{Journal of \LaTeX\ Class Files,~Vol.~14, No.~8, August~2021}%
{Shell \MakeLowercase{\textit{et al.}}: A Sample Article Using IEEEtran.cls for IEEE Journals}


\maketitle

\begin{abstract}
In daily life, we encounter a variety of sounds, both desirable and undesirable, with limited control over their presence and volume. Our work introduces “Listen, Chat, and Remix” (LCR), a novel multimodal sound remixer that controls each sound source in a mixture based on user-provided text instructions. LCR distinguishes itself with a user-friendly text interface and its unique ability to remix multiple sound sources simultaneously within a mixture, without needing to separate them. Users input open-vocabulary text prompts, which are interpreted by a large language model to create a semantic filter for remixing the sound mixture. The system then decomposes the mixture into its components, applies the semantic filter, and reassembles filtered components back to the desired output. We developed a 160-hour dataset with over 100k mixtures, including speech and various audio sources, along with text prompts for diverse remixing tasks including extraction, removal, and volume control of single or multiple sources. Our experiments demonstrate significant improvements in signal quality across all remixing tasks and robust performance in zero-shot scenarios with varying numbers and types of sound sources. An audio demo is available at: https://listenchatremix.github.io/demo.
\end{abstract}


\section{Introduction}

Imagine yourself at a cocktail party, listening to a conversation between two people.  Amidst their conversation, a soft guitar melody gracefully wafts through the air, however woefully accompanied by the annoying noises of passing cars outside. At this moment, you open the smart sound remixer and type \textit{``Hey. Can you reduce the volume of the excited male speaker and remove the background traffic noise completely?"} Instantly, you find it much easier to follow their conversation, and the soothing music remains hovering in the air. 

The scenario where people want to focus on specific sounds among multiple concurrent ones is known as the cocktail party problem \cite{cherry1953some, haykin2005cocktail, mcdermott2009cocktail}. Traditional methods for enhancing the hearing experience in multi-sound environments\cite{kates2008digital, clark2014technology, launer2016hearing} usually amplify or suppress a wide spectrum of sounds within a mix, but they fall short in targeting specific sound sources. Some recent models can extract a target sound source conditioned on neural auditory attention signals \cite{van2016eeg, han2019speaker} or sound word labels \cite{kilgour2022text, liu22w_interspeech, semantic_hearing}. However, these models concentrate on isolating a single source and lack the versatility to control multiple sound sources. Moreover, most of these models do not provide an interface that allows users to easily modify the soundscape based on textual instructions.

\begin{figure}[!t]
\centering
\includegraphics[width=0.9\columnwidth]{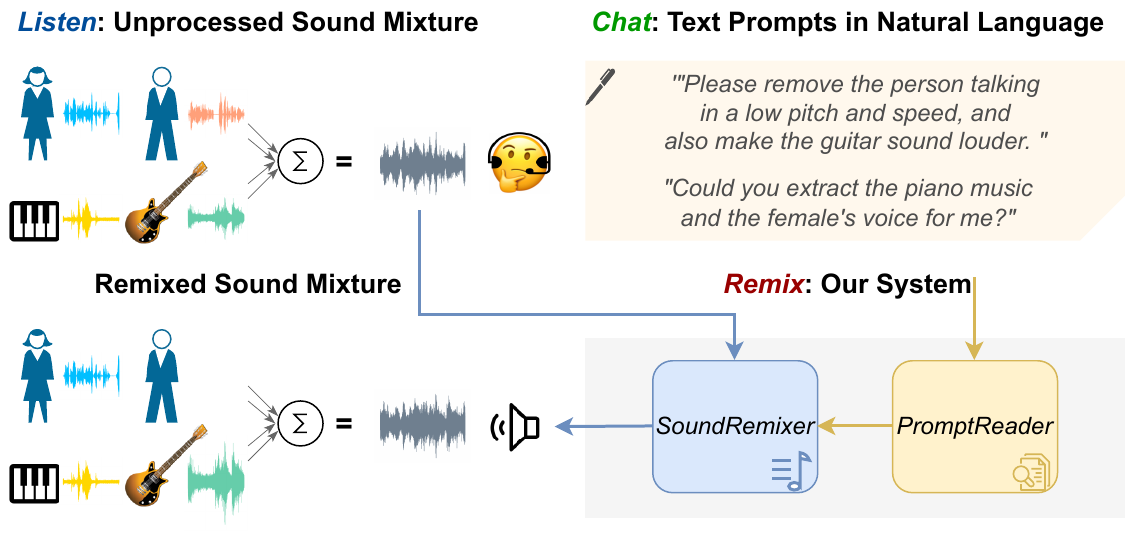}
\caption{An overview of \textit{Listen, Chat, and Remix}.}
\label{fig:overview}
\vspace{-0.6cm}
\end{figure}

In this work, we study text-guided soundscape remixing, a new problem that aims to simultaneously extract, remove, or control the volume of multiple sounds in a mixture based on the user's open-vocabulary instruction. We propose the first text-guided sound remixer: \textit{Listen, Chat, and Remix} (LCR). Figure \ref{fig:overview} shows the schematic of LCR: To begin with, a user and LCR both \textit{Listen} to the sound mixture. Next, the user \textit{Chat}s with a text prompt in natural language to specify which sound sources to be remixed and in which manner. Finally, LCR \textit{Remix}s the sound mixture according to the text prompt and outputs a new sound mixture that aligns with the user's command. LCR differs from previous works in three distinctive features. 1) \textbf{Selectivity}: LCR recognizes and selects target talkers by semantic descriptions. A user describes speaking styles for human speakers, including gender, pitch, tempo, energy, and emotion, and the category for non-speech sounds, such as the name of an animal or a musical instrument. 2) \textbf{Accessibility}: LCR employs a large language model (LLM) to interpret users' open-vocabulary text prompts for different remixing tasks. This textual instruction is more user-friendly than other forms of instruction like neural signals, speaker features, sound labels, and so on. 3) \textbf{Soundscape remixing}: LCR simultaneously remixes multiple sound sources in the sound mixture in one step, without separating them in advance and scaling each sound one by one. Therefore, LCR does not require clean sources for training.

We evaluate LCR on all combinations of extraction, removal, and volume control for one or more sound sources in the mixture. Since the sound mixtures and text prompts required to train LCR did not exist, we have curated a sound mixture dataset, comprising around 160 hours or over 100k of speech and audio mixtures with 500k of text prompts written for different remixing tasks. Our experiments show that LCR trained on this dataset enhances the signal quality in around 94\% of sound mixtures and improves the signal-to-noise ratio by 10.4 or 11.4 dB averaged across all remixing tasks for our transformer \cite{transformer} or Mamba \cite{mamba} models. Notably, LCR surpasses previous expert models on target speech or audio extraction as one of the remixing tasks in both objective and subjective evaluation. Moreover, LCR can perform zero-shot remixing on both synthetic and real sound mixtures containing unseen types of sounds or different numbers of sources. 

\section{Related Works}

\textbf{Sound Separation and Extraction} Soundscape understanding and processing \cite{settle2018end, turpault2020improving} relies on sound separation and extraction, which provides the computational groundwork to isolate each sound source in a mixture. Sound separation models can separate speeches \cite{deep_clustering, convtasnet, sepformer}, music \cite{bryan2013source, demucs, bandsplitrnn}, and universal sounds \cite{uss, fuss, cocktailfork}. However, they are not selective since they unconditionally separate all sources in the mixture. On the other hand, sound extraction models selectively extract one target sound from a sound mixture \cite{tse_survey2, tse_survey1}. A clue, which can take multiple forms, is provided to identify the target. Some common clues include lip movement videos \cite{afouras2018conversation, av_conformer}, videos recording the sound production \cite{PixelPlayer, AudioScope}, target speaker embeddings \cite{SpeakerBeam, VoiceFilter}, neural auditory attention signals \cite{van2016eeg, han2019speaker}, locations \cite{Gu2019NeuralSF, Heitkaemper2019ASO}, the language spoken \cite{tle}, and recently emerging text prompts. 

\textbf{Text-Guided Audio Applications} Recent advancements in NLP and text-audio learning have led to the use of text as guidance for various audio applications. Target sound extraction models have used semantic labels, descriptions of audio sources \cite {kilgour2022text, liu22w_interspeech, liu2023separate, semantic_hearing}, transcriptions, or speaker attributes \cite{hao2023typing} as new forms of clues. Text prompts have also been introduced in sound generation \cite{TextrolSpeech, audiogen, audioldm} and editing \cite{audit, usee, Voicebox, audiobox} as well. This work focuses on text-guided soundscape remixing, a new category of text-guided audio applications. The problem we address is most similar to target sound extraction, which is one of the many tasks that LCR can solve. Although the problem is unique, we have curated the instruction dataset in a similar manner to \cite{TextrolSpeech, instructme}, which generate the language descriptions or instructions of sounds from their metadata.

\textbf{Text-Audio Interface} To execute the correct audio processing task from text instructions, a shared interface between text and audio modalities is necessary. Two standard designs exist for this interface. Composite systems like AudioGPT \cite{audiogpt} and WavCraft \cite{wavcraft} use an LLM to analyze the user's prompt and generate an executable instruction that calls downstream audio models. These systems can handle as many tasks as the connected audio models support, with little or no finetuning. However, they are often slow due to the extra time needed to decode and execute the instruction and can be error-prone if the LLM outputs an instruction the audio model doesn't recognize. Conversely, joint-modeling systems enhance speed and performance by using text embeddings (without decoding) as the interface and finetuning the LLM with audio data. Some systems for sound extraction \cite{liu2023separate} and audio generation \cite{audioldm} utilize text embeddings from Contrastive Language-Audio Pretraining (CLAP) \cite{CLAP2022, CLAP2023}, which aligns text and audio embeddings. Other joint-modeling systems for sound extraction \cite{kilgour2022text, hao2023typing} finetune the LLM and audio model together with a task-specific objective in an end-to-end manner. In this work, we optimize LCR in an end-to-end manner as the latter.

\section{soundscape remixing} \label{sec:sm2me}
\begin{figure*}[!t]
  \centering
\includegraphics[width=\textwidth]{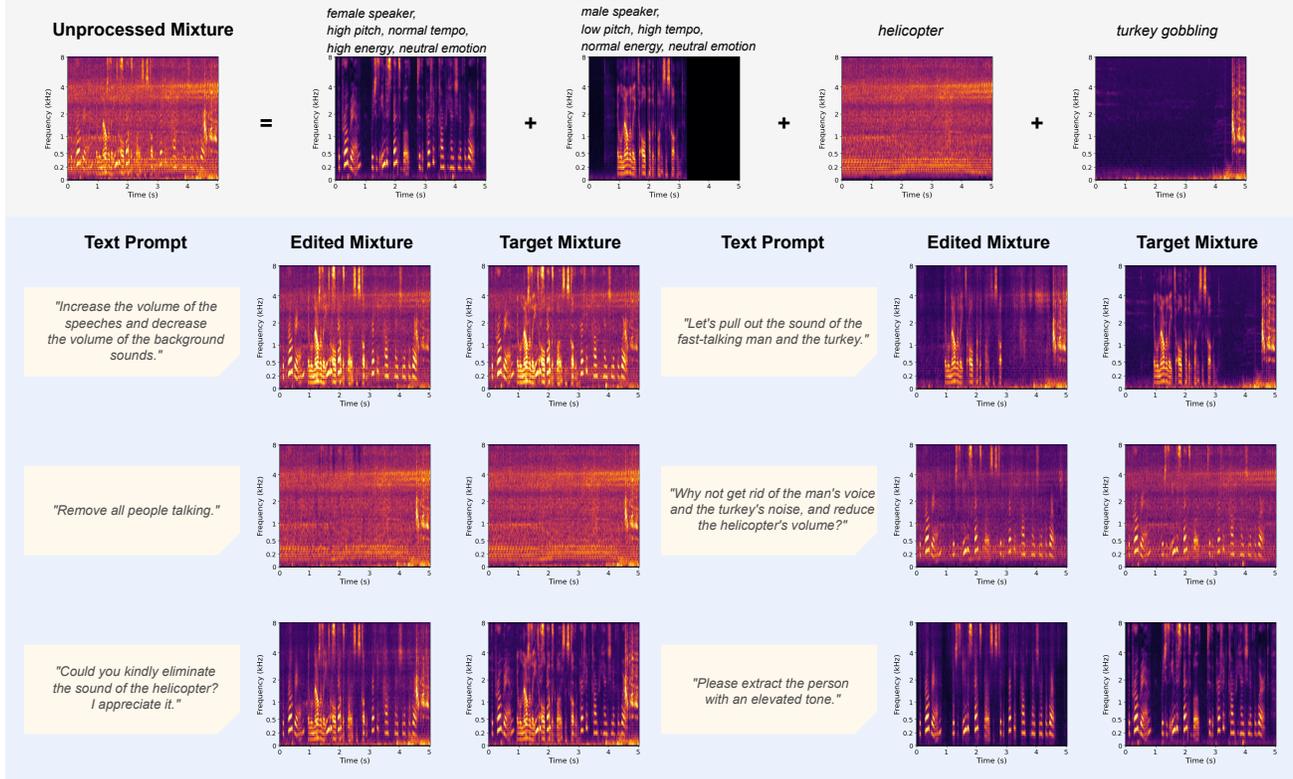}
  \caption{An example sound mixture consists of a female speaker, a male speaker, a helicopter, and a turkey. We wrote 6 example prompts covering S$\uparrow$, SR, TAR, ME, MEVC, and TSE task. The unprocessed, remixed, and the target Mel spectrograms are plotted for comparison.}
  \label{fig:spec1}
\end{figure*}

We formulate soundscape remixing in this section. Figure \ref{fig:spec1} provides an example of such remixing. We show six example text prompts that users can input to LCR, including adjusting the presence and volume of one or more speakers or audio sources. According to the text prompt, LCR remixes the sound mixture differently, as shown in six different spectrograms.

\subsection{Sources and Actions}
\label{sec:source_and_action}
Consider a sound mixture $x$, we can express it as the sum of $N$ different sources: $x = \sum_{i=1}^{N} s_i \label{eq:input_mixture}$.
Each source can have different energy and can be either a human speech or a non-speech sound: $s_i \in \mathcal{S} = \mathcal{S}_{\text{speech}} \cup \mathcal{S}_{\text{audio}}$. For the remainder of this paper, we will use the term $\textit{audio}$ for non-speech sounds, including animal voices, sound effects, music, and noises. We can describe each sound source $s_i$ by a semantic description $\tilde{s}_i$, which can be the speaking style measured in one or multiple of gender, pitch, tempo, energy, and emotion for a speech source or the class label for an audio source. Let us define $\tilde{\mathcal{S}}$ as the set of all semantic descriptions. In this work, we assume that no two sound sources share the same semantic description in the sound mixture. In other words, no two speech sources share the same speaking style, and no two audio sources share the same class label. Therefore, every sound mixture is a combination of $N$ different sources of different descriptions $\{\tilde{s}_1, \tilde{s}_2, ..., \tilde{s}_N\}$, with $\tilde{s}_i \neq \tilde{s}_j, \forall i \neq j\large$, and a description $\tilde{s}_i$ can identify an unique source $s_i$ in the mixture.

\begin{table*}[ht]
\caption{All remixes grouped into 16 tasks by \textit{One speech, one audio, or multiple?} \textit{Extract, remove, or control the volume?}}
\label{tab:tasks}
\begin{center}
\begin{sc}
\begin{adjustbox}{width=\textwidth,center}
\begin{tabular}{ccccc}
\toprule
One Speech & One Audio & All Speeches or Audios & Multiple Speeches and Audios & All Sounds \\
\hline
\textit{Target Speech Extraction (TSE)} & \textit{Target Audio Extraction (TAE)} & 
\textit{Speech Enhancement (SE)} & \multirow{4}{*}{\makecell{\textit{Multiple Sound Extraction or Removal (ME)} \\ \textit{Mult. Sound Ext. or Rem. and Vol. Ctrl. (MEVC)} \\ \textit{Multiple Sound Volume Control (MVC)}}} & \\
\textit{Target Speech Removal (TSR)} & \textit{Target Audio Removal (TAR)} & 
\textit{Speech Removal (SR)} & & \\
\textit{Target Speech Volume Up (TS$\uparrow$)} & \textit{Target Audio Volume Up (TA$\uparrow$)} & 
\textit{Speech Volume Up (S$\uparrow$)} &  & \hspace{-0.4cm}\multirow{2}{*}{\textit{Overall Volume Control (OVC)}}
 \\
\textit{Target Speech Volume Down (TS$\downarrow$)} & \textit{Target Audio Volume Down (TA$\downarrow$)} & 
\textit{Speech Volume Down (S$\downarrow$)} & &
 \\
\bottomrule
\end{tabular}
\end{adjustbox}
\end{sc}
\end{center}
\end{table*}

Our goal is to remix the mixture $x$ to yield a new mixture $y$, which is made up of the same sources with the same or different scaling factors: $y = \sum_{i=1}^{N} \alpha_i s_i$. The values of $\alpha_i$ are determined by the user's desired remix, which is a combination of actions in $\mathcal{A} = \{\textit{removing} \squad (0), \textit{keeping} \squad (1), \textit{increasing the volume} \squad (\uparrow), \textit{decreasing the volume} \squad (\downarrow)\}$. These actions correspond to scaling factors $\mathcal{A}_\alpha = \{0, 1, 2, 0.5\}$. For example, a scaling factor $\alpha_i = 2$ implies increasing the volume of $s_i$ by 6 dB. If only one $\alpha_i$ is equal to 1, and all others are equal to 0, it instructs the extraction of $s_i$ or the removal of all other sources, both resulting in $y=s_i$.

\subsection{Remixing Tasks} \label{sec:remixing_task}
A tuple $(a_i, \tilde{s}_i) \in \mathcal{A} \times \tilde{\mathcal{S}}$ specifies that the user wants to do $a_i$ for a source $\tilde{s}_i$ in the mixture. Therefore, for a mixture of $N$ sources, a collection of $N$ such tuples: $\pi = \{(a_1, \tilde{s}_1), ..., (a_N, \tilde{s}_N)\}$ specifies how the remixed mixture should sound like. We call $\pi$ a remixing instruction. If the composition $\{\tilde{s}_1, \tilde{s}_2, ..., \tilde{s}_N\}$ of the sound mixture is fixed, $\mathcal{A}^{N}$ represents the set of all possible remixes to this mixture. For instance, if a mixture contains four sources, there is a total of $|\mathcal{A}|^4-2 = 254$ possible remixes, excluding the trivial identity and silence cases. One example remixing $[0, \downarrow, \uparrow, 1] \in \mathcal{A}^{4}$ instructs the following: remove $s_1$, turn down the volume of $s_2$, turn up the volume of $s_3$, and keep $s_4$.

Notice that many elements in $\mathcal{A}^{N}$ are similar. For example, $[0, \downarrow, \uparrow, 1]$ and $[\downarrow, 0, 1, \uparrow]$ specify same actions but applied to different sources. To more accurately evaluate the performance of LCR and to maintain consistency with existing literature, we categorize $\mathcal{A}^{N}$ into the task space $\mathcal{T}$. For any sound mixture containing at least two speech sources and at least two audio sources, there are 16 different tasks in $\mathcal{T}$. We present $\mathcal{T}$ based on which source(s) are mixed (in columns) and the manner in which they are mixed (in rows) in Table  \ref{tab:tasks}. Note that $\mathcal{T}$ is a super set of the target speech or audio extraction and removal tasks widely studied in previous works \cite{semantic_hearing, liu2023separate, hao2023typing}.

\subsection{Text Prompts}
Although $\pi = \{(a_1, \tilde{s}_1), ..., (a_N, \tilde{s}_N)\}$ itself accurately specifies a remixing instruction, writing an instruction in this format can be inconvenient and difficult for humans, because we prefer to use language rather than numbers and symbols. As a result, we adopted $p = \mathcal{H} (\pi)$ as the input for LCR. Here, $p$ is an open-vocabulary text prompt written in natural language by $\mathcal{H}$, which can be a human or a large language model (LLM). Given the impracticality of hiring humans to write hundreds of thousands of prompts, we leveraged a LLM to generate text prompts $p$ from $\pi$. The text prompt generation pipeline can be found in Section \ref{sec:dataset}. 

\begin{figure*}[ht]
\centering
\includegraphics[width=0.95\textwidth]{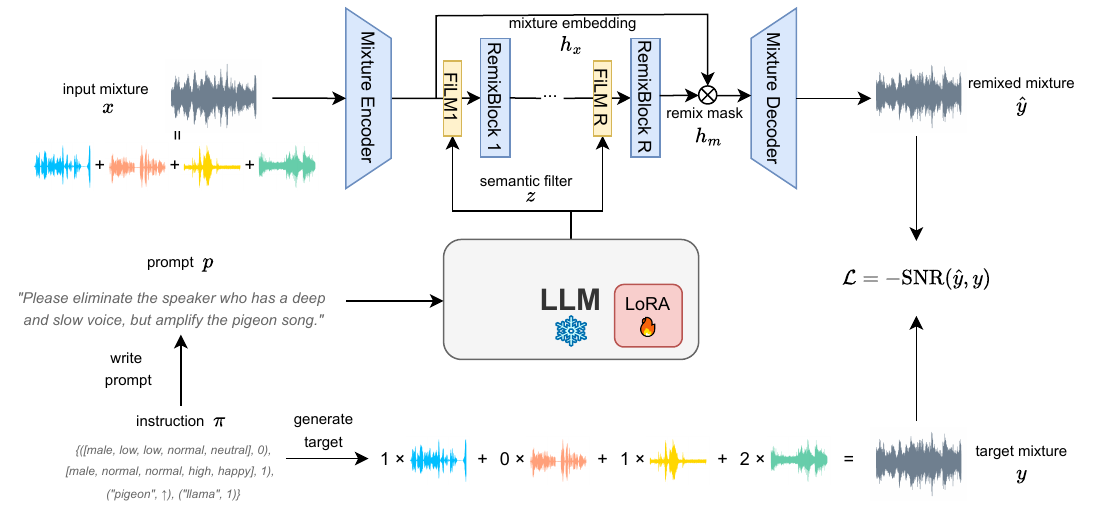}
\caption{The components and the training paradigm of LCR.}
\label{fig:system}
\end{figure*}

\section{Listen, Chat, and Remix}
Listen, Chat, and Remix (LCR) is our solution to the soundscape remixing problem. LCR remixes a sound mixture $x$ given a natural language text prompt $p$: $\hat{y} = \text{LCR}(x, p)$. LCR includes two models: a \textit{PromptReader} to read text prompts and a \textit{SoundRemixer} to remix sound mixtures. A graphical illustration of it in details is depicted in Figure \ref{fig:system}.

\subsection{Prompt Reader}

The \textit{PromptReader} is a language model that behaves as the inverse of $\mathcal{H}$: It generates a $D$-dimensional text embedding $z \in \mathbb{R}^D$ from a text prompt $p$, which encodes the information about the sources and actions specified in  $\pi$. We call $z$ a \textit{semantic filter} because it is generated from text to guide the \textit{SoundRemixer} to execute proper filtering of sounds. We show the correspondence between the semantic filter and the resulting acoustic filter in Section \ref{sec:analysis}.
We employed a pretrained language model GPT-2 \cite{gpt2} or LLaMA2 \cite{llama2} as the \textit{PromptReader}. We finetuned it, with full parameters or low-rank approximation \cite{hu2022lora}, using the gradients backpropagated from the \textit{SoundRemixer}. While the pretrained language model only grasps textual information, further joint training with \textit{SoundRemixer} enables it to generate text embeddings that exhibit awareness of the acoustic characteristics (e.g., pitch, emotion, animal voices, and instrument timbres) written in the descriptions (see Section \ref{sec:analysis}). Consequently, the enhanced embeddings facilitate the \textit{SoundRemixer} in more easily identifying the corresponding sources mentioned in the text prompt.

\begin{figure}[h]
\centering
\setlength{\fboxsep}{0pt} 
\setlength{\fboxrule}{1pt} 
\includegraphics[width=\columnwidth]{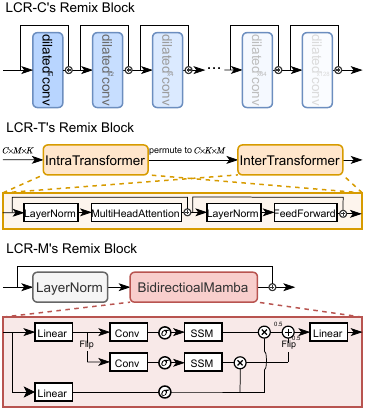}
\caption{A closer look at three different kinds of remix blocks for LCR-C,  LCR-T, and LCR-M.}
\label{fig:blocks}
\end{figure}

\subsection{Sound Remixer}
The \textit{SoundRemixer} is an acoustic model responsible for remixing the sound mixture $x$ into $\hat{y}$ based on the semantic filter $z$.
We adopted a similar architecture of source separation models for the \textit{SoundRemixer} due to their ability to isolate individual sources in a sound mixture. However, it is essential to clarify that \textit{SoundRemixer} does not explicitly (at any stage) separate all the sources in the mixture to remix them. Instead, the model functions similarly to a sound extraction model, directly producing a single sound that is a remix of all sources.

\textit{SoundRemixer} is a time-domain model with a pair of learnable linear encoder $\mathcal{E}: \mathbb{R}^{1 \times T} \rightarrow \mathbb{R}^{C \times L}$ for mapping the input waveform $x$ of $T$ samples into a $C$-dimensional latent representation $h_x$ and a learnable linear decoder $\mathcal{D}: \mathbb{R}^{C \times L} \rightarrow \mathbb{R}^{1 \times T}$ for mapping the latent $h_{\hat{y}}$ back to the waveform $\hat{y}$:
\begin{align}
    h_x = \mathcal{E}(x), \quad
    \hat{y} = \mathcal{D}(h_{\hat{y}})
\end{align}
$\mathcal{E}$ and $\mathcal{D}$ are implemented by one-dimensional convolution and transpose convolution, respectively, both with a kernel size of 16 samples (1 millisecond) and a stride $\floor{T / L}$ of 8. In between, a mask network denoted as $\mathcal{M}$ estimates a single remixing mask $h_m \in \mathbb{R}^{C \times L}$ that selectively extracts, removes, amplifies, or reduces samples in $h_x$ according to $z$ through element-wise multiplication on $h_x$:
\begin{align} \label{eq:mask}
    h_m = \mathcal{M}(x, z), \quad
    h_{\hat{y}} = h_m \otimes h_x
\end{align}
$\mathcal{M}$ consists of $R$ remix blocks with the same structure. We have experimented with three different blocks: temporal convolutional network (TCN) \cite{TCN}, dual-path (dp) transformer \cite{transformer}, and bidirectional Mamba \cite{bimamba}, inspired by Conv-TasNet \cite{convtasnet}, Sepformer \cite{sepformer}, and Mamba-TasNet \cite{mambatasnet}, respectively. We call LCRs implemented by convolutional, transformer, or Mamba blocks \textbf{LCR-C}, \textbf{LCR-T}, and \textbf{LCR-M}. The dimensions of the encoder, $C$, the same as the dimensions that remix blocks operate on, are 512 for LCR-C and 256 for LCR-T and LCR-M, following \cite{convtasnet, sepformer, mambatasnet}.

Figure \ref{fig:blocks} provides a detailed breakdown of each type of remix block. The remix block in LCR-C is a TCN of eight depthwise convolution layers \cite{depthwise_conv}, where the dilation factor increases exponentially from $2^0$ to $2^7$. LCR-C incorporates three such TCNs in total. Next, LCR-T adopts a dual-path architecture to divide the sound sequence into smaller chunks to prevent the quadratic complexity of self-attention. We use a chunk size $K$ of 250 samples with an overlap of 125 samples, resulting in $M$ chunks. Each remix block contains an \textit{IntraTransformer} module to process the $K$ samples within each chunk and an \textit{InterTransformer} module to handle all $M$ chunks together, with both modules containing eight transformer layers. Lastly, LCR-M simply stacks 32 bidirectional Mamba layers \cite{bimamba} (i.e., single-path, with the same number of layers as LCR-T) thanks to the linear complexity of Mamba.

For all kinds of the remix blocks, we guide them by the semantic filter $z$ using Feature-wise Linear Modulation (FiLM) \cite{film}. Before the $i$th block, the text-guided feature $\tilde{h}_i$ is computed by:
\begin{align}
    &\tilde{h}_i = \gamma_i h_i + \beta_i, \quad 0 \leq i \leq R-1 \\
    &\gamma_i = \mathcal{F}_i(z), \quad \beta_i = \mathcal{G}_i(z)
\end{align}
$\mathcal{F}_i: \mathbb{R}^{D} \rightarrow \mathbb{R}^{C}$ and $\mathcal{G}_i: \mathbb{R}^{D} \rightarrow \mathbb{R}^{C}$ are two two-layer perceptrons which project the $z$ to the same dimension of $h_i$. Afterwards, $\gamma_i$ and $\beta_i$ are replicated along the time axis to modulate $h_i$. This ensures that the text prompt maintains time-invariant influence over the duration of the mixture signal.

\subsection{Training Objective}
Each training instance is a tuple $(x, y, p)$ including an input mixture $x$, a target mixture $y$, and a text prompt $p$. The sources in the mixture $\{s_1, s_2, ..., s_N\}$ and the underlying symbolic instruction $\pi$ used in the creation of $y$ and $p$ are hidden from the system. LCR estimates a mixture $\hat{y}$ given $x$ and $p$. All model components are jointly optimized by maximizing the signal-to-noise ratio (SNR) between the target and the estimated mixture.
Since LCR does not require clean sources $\{s_1, s_2, ..., s_N\}$ for training, it does not suffer from the $\mathcal{O}(N!)$ complexity in calculating the source-level permutation-invariant training (PIT) \cite{pit} loss (or $\mathcal{O}(N^3)$ with Hungarian algorithm \cite{hungarian_pit}). Therefore, LCR is scalable to the number of the sources with a constant training complexity.

\section{Dataset} \label{sec:dataset}
Due to the absence of natural language instruction-prompted sound remixing datasets, we have generated our own sound mixtures and text prompts. We highlight key features of our dataset and will provide more details when it is open-sourced. 

\noindent
\textbf{1) Sound Sources} 
Speech sources were from TextrolSpeech \cite{TextrolSpeech} and audio sources were from VGGSound \cite{Vggsound} (for training and in-domain evaluation) or FSD50K \cite{fsd50k} (for zero-shot evaluation). Speakers and audios were split into training, validation, and testing sets. All sources were sampled to 16 kHz and cropped or padded to 5 seconds (partially overlapped).

\noindent
\textbf{2) Sound Mixtures}
We generated 100k, 5k, and 10k mixtures of 2 speech + 2 audio sources from TextrolSpeech and VGGSound for training, validation, and testing. These four-source mixtures, featuring a variety of categories, simulate acoustic scenes commonly encountered in daily life. We also generated 5,000 mixtures of 2 speech, 2 audio, 2 speech + 1 audio, 1 speech + 2 audio for zero-shot evaluation on unseen numbers of sources, and 5,000 mixtures of 2 speech + 2 audio sources from FSD50K with either all classes or unseen classes for zero-shot evaluation on unseen sounds.

\noindent
\textbf{3) What \& How to Remix} For every mixture, we first evenly sampled a task $t \in \mathcal{T}$ (e.g. target speech extraction), and then sampled a particular instruction $\pi$ of task $t$ (e.g. extracting a particular speaker). The reported performance was calculated by averaging the performance of all mixtures of that task.

\noindent
\textbf{4) Text Prompts} We requested GPT-3.5 Turbo to write a text prompt five times from the assigned instruction $\pi$ for each mixture. To improve the diversity of the text prompts, we asked it to try imperative or interrogative sentences and substitute synonyms for class labels and style keywords such as \textit{female}, \textit{male}, \textit{low}, \textit{high}, \textit{pitch}, and \textit{tempo}. 

\begin{table*}[!t]
\caption{Remixing performance (SNRi in dB) of LCR-\{C, T, M\} and baseline models for extraction (E), removal (R), or volume control (VC), up ($\uparrow$) and down ($\downarrow$), applied to target (T) speech (S), audio (A), or the overall (O) mixture. The best scores are shown in bold for each task.}
\label{tab:all}
\begin{sc}
\begin{adjustbox}{width=0.9\textwidth,center}
\begin{tabular}{ccccccccccc}
\toprule
\textit{SoundRemixer} & \textit{PromptReader} & TSE & TSR & TS$\uparrow$ & TS$\downarrow$ & TAE & TAR & TA$\uparrow$ & TA$\downarrow$ & SE \\
\midrule
\multicolumn{2}{c}{\textbf{\textit{Ideal Cascaded System}}} & & & & & & & & &\\
\multirow{2}{*}{Conv-TasNet (PIT)} & \textit{PIT + G.T. Actions} & 13.3 & 7.7 & 7.5 & 6.9 & 10.1 & 4.2 & 4.1 & 3.8 & 9.6 \\
& 2 CLAPs \textit{+ G.T. Actions} & 12.3 & 6.6 & 6.3 & 6.0 & 9.0 & 3.0 & 3.0 & 2.6 & 9.6 \\
\midrule
\multicolumn{2}{c}{\textbf{\textit{Reproduced LASS \cite{liu2023separate}}}} & & & & & & & & &\\
\multirow{2}{*}{TCN} & CLAP (Speech) & 8.8 & - & - & - & - & - & - & - & - \\
& CLAP (Audio) & - & - & - & - & 8.1 & - & - & - & - \\
\midrule
\multicolumn{2}{c}{\textbf{\textit{Listen, Chat, and Remix}}} & & & & & & & & &\\
\multirow{3}{*}{TCN} & GPT-2 (finetuned) & 8.0 & 3.1 & 3.4 & 3.7 & 7.5 & 1.6 & 2.6 & 2.4 & 7.8 \\
 & LLaMA2 (frozen) & 7.0 & 2.2 & 2.7 & 2.7 & 7.1 & 1.3 & 2.0 & 1.9 & 5.1 \\
 & LLaMA2 (LoRA) & 9.7 & 4.6 & 4.7 & 4.8 & 8.2 & 2.5 & 3.1 & 3.0 & 8.2 \\
\multirow{2}{*}{DP Transformer} & GPT-2 (finetuned) & 11.1 & 6.1 & 6.5 & 6.9 & 8.8 & 3.2 & 4.1 & 4.0 & 9.7 \\
 & LLaMA2 (LoRA) & 13.1 & 7.9 & 8.1 & 8.3 & 10.1 & \textbf{4.5} & 4.8 & 4.8 & 10.7 \\
 \textbf{Bidirectional Mamba} & \textbf{LLaMA (LoRA)} & \textbf{13.7} & \textbf{8.4} & \textbf{8.8} & \textbf{9.1} & \textbf{10.3} & \textbf{4.5} & \textbf{5.1} & \textbf{5.0} & \textbf{11.7} \\
\bottomrule
\multicolumn{11}{c}{Continued} \\
\toprule
\textit{SoundRemixer} & 
\textit{PromptReader} & SR & S$\uparrow$ & S$\downarrow$ & ME & MVC & MEVC & OVC & \multicolumn{2}{c}{Average} \\
\midrule
\multicolumn{2}{c}{\textbf{\textit{Ideal Cascaded System}}} & & & & & & & & &\\
\multirow{2}{*}{Conv-TasNet (PIT)} & \textit{PIT + G.T. Actions} & 9.9 & 8.6 & 8.4 & \textbf{6.9} & 6.2 & 7.1 & 21.7 & \multicolumn{2}{c}{8.5} \\
& 2 CLAPs \textit{+ G.T. Actions} & 9.9 & 8.6 & 8.4 & 5.3 & 4.9 & 6.1 & 21.7 & \multicolumn{2}{c}{7.7} \\
\midrule
\multicolumn{2}{c}{\textbf{\textit{Listen, Chat, and Remix}}} & & & & & & & & &\\
\multirow{3}{*}{TCN} & GPT-2 (finetuned) & 8.0 & 6.4 & 6.0 & 3.4 & 2.6 & 3.3 & 42.7 & \multicolumn{2}{c}{7.0} \\
 & LLaMA2 (frozen) & 6.8 & 5.0 & 5.0 & 2.8 & 1.3 & 2.0 & 39.5 & \multicolumn{2}{c}{5.9} \\
 & LLaMA2 (LoRA) & 8.5 & 7.4 & 7.0 & 4.2 & 3.9 & 4.6 & 40.3 & \multicolumn{2}{c}{7.8} \\
\multirow{2}{*}{DP Transformer} & GPT-2 (finetuned) & 10.0 & 9.1 & 8.6 & 5.2 & 4.5 & 5.0 & 43.8 & \multicolumn{2}{c}{9.2}  \\
 & LLaMA2 (LoRA) & 11.1 & 10.2 & 9.6 & 6.5 & 6.3 & 6.8 & 44.2 & \multicolumn{2}{c}{10.4} \\
\textbf{Bidirectional Mamba} & \textbf{LLaMA (LoRA)} & \textbf{12.0} & \textbf{11.1} & \textbf{10.6} & 6.5 & \textbf{6.6} & \textbf{7.2} & \textbf{52.1} & \multicolumn{2}{c}{\textbf{11.4}} \\
\bottomrule
\end{tabular}
\end{adjustbox}
\end{sc}
\end{table*}

\section{Experiments}

We evaluated LCR by 16 remixing tasks in Table \ref{tab:tasks}. As LCR is the first model capable of reading text prompts and performing all 16 tasks, we compared it with cascaded sound separation systems, the text-guided target audio extraction model AudioSep \cite{liu22w_interspeech, liu2023separate}, the speech extraction model LLM-TTS \cite{hao2023typing}, and Sepformer \cite{sepformer} trained for speech enhancement \cite{dubey2022icassp}. Objective metrics are compared in Tables \ref{tab:all} to \ref{tab:tencent}, with human evaluation results in Table \ref{tab:audioset_zs}. In addition, we conducted an in-depth analysis of \textbf{LCR-T}(ransformer) and \textbf{LCR-M}(amba) across all remixing tasks and zero-shot evaluation on mixtures with unseen numbers and types of sounds and mixtures in the wild. We further examined the impact of speaking style on speech remixing. For ablations on alternative model configurations, including a causal \textbf{LCR-C}(onvolution) model, and on text prompts, see Section \ref{sec:ablations}. Analyses of semantic and acoustic features are presented in Section \ref{sec:analysis}. Details on the model and training are in Appendix \ref{app:training}.

\subsection{Baseline Systems and Model-wise Performance Analysis} \label{sec:bs_mwpa}

We compared LCR with two baseline models. One is ideal cascaded systems with ground truth remixing scales and the other is language-queried audio source separation (LASS) models \cite{liu22w_interspeech, liu2023separate}. We evaluated the remixing performance by SNR improvement (SNRi) with the target sound mixtures. The cascaded systems separate all sound sources, scale each source individually, and sum them up. The first cascaded system finds the target sources by minimizing PIT loss \cite{pit} and remixes them with ground truth actions. The second cascaded system matches the target sources with two CLAP models (audio and speech\footnote{We used the official CLAP checkpoint \cite{CLAP2022} for audio, and trained another CLAP on TextrolSpeech for speech. Same CLAPs are used for LASS models.}) and also remixes them with ground truths. LCR lags behind these ideal systems slightly because they perfectly interpret both actions and sources or at least the actions with ground truths. However, LCR still shows competitive performance compared to the ideal systems, and LCR with a stronger \textit{PromptReader} LLaMA2 finetuned can even surpass the second system in three tasks (TA$\uparrow$, TA$\downarrow$, and OVC). In another comparison with LASS for target speech or audio extraction, LCR with finetuned LLaMA2 also outperforms expert extraction models and can handle 14 more tasks. An even stronger performance is observed with both stronger \textit{SoundRemixer} Transformer or Mamba and \textit{PromptReader} LLaMA2, compared to their weaker counterparts TCN and GPT-2. LCR-M, Mamba + finetuned LLaMA2, performs the best in all tasks with an average SNRi of 11.4 dB, and LCR-T with transformer follows with an average SNRi of 10.4 dB.

Finetuning \textit{PromptReader} is also necessary. This improves all tasks by 2 dB compared to freezing LLaMA2.

\begin{figure}[!ht]
\centering
\includegraphics[width=\columnwidth]{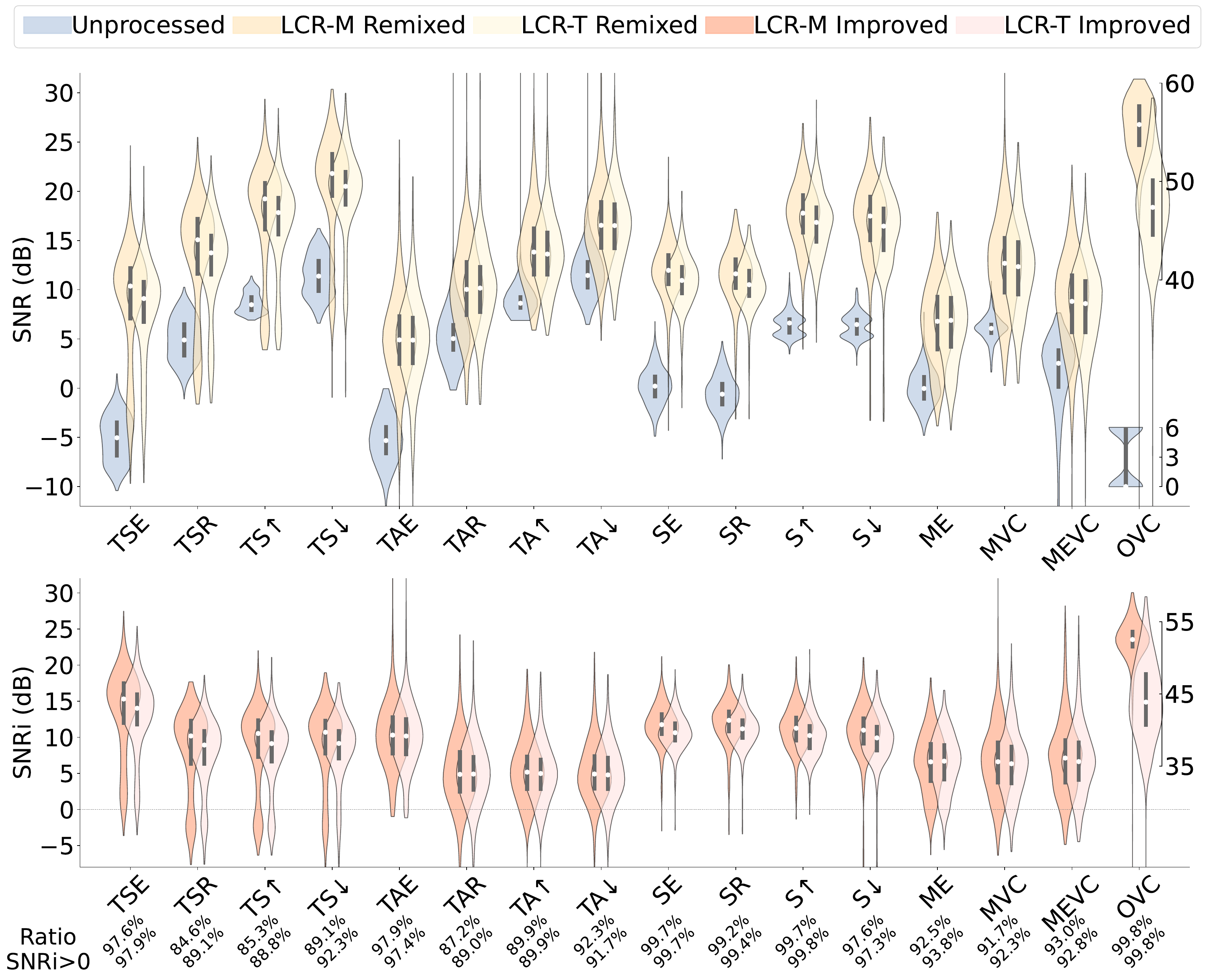}
\caption{The SNR distribution (25th, 50th, and 75th quartile) of the unprocessed (blue) and LCR-M or LCR-T remixed (darker or lighter orange) sound mixtures and their differences (SNRi, darker or lighter pink), calculated with respect to the target mixtures. The ratios of mixtures with an improvement (SNRi $> 0$) for each task and model are written below the task labels.}
\label{fig:snr}
\end{figure}

\begin{figure*}[!ht]
\centering
\setlength{\fboxsep}{0pt} 
\setlength{\fboxrule}{1pt} 
\includegraphics[width=\textwidth]{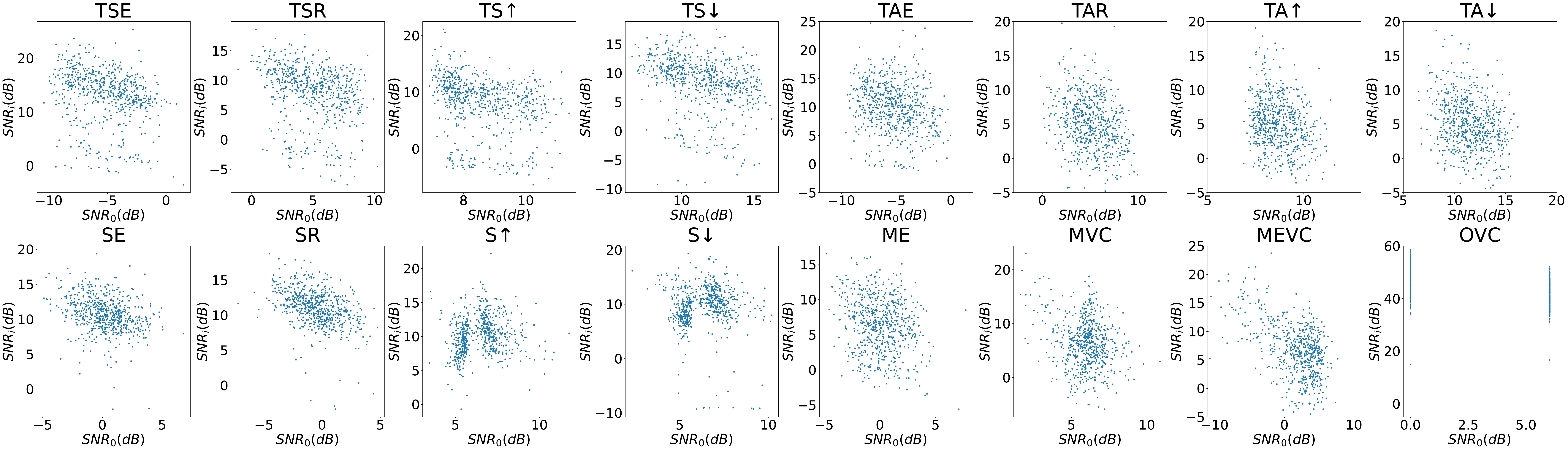}
\caption{The SNR improvement of LCR-T's remixed mixtures (SNRi, y-axis) vs. the initial SNR of the unprocessed mixtures (SNR$_{0}$ x-axis), with respect to the target mixture. Each subplot corresponds to one of the 16 tasks.}
\label{fig:snr_xy}
\end{figure*}

\subsection{Task-wise Performance Analysis}

We analyzed the remixing performance of the top-performing LCR-M and LCR-T across all 16 tasks. We reported the average SNRi of each task in Table \ref{tab:all}. The difficulty of tasks varies significantly. Overall volume control (OVC), target speech/audio extraction/removal (TSE, TSR, TAE, TAR), and speech enhancement/removal (SE, SR) perform better, with an average SNRi over 10 dB. These tasks are also more commonly studied in the literature. In contrast, target speech/audio removal (TSR, TAR) or volume control (TS$\uparrow$, TS$\downarrow$, TA$\uparrow$, TA$\downarrow$) and multiple sound extraction or/and volume control (MVE, MVC, MEVC) are more challenging, which are also less studied. We also observe that LCR performs on average 3 dB worse on audio tasks (TAE, TAR, TA$\uparrow$, TA$\downarrow$) than speech tasks (TSE, TSR, TS$\uparrow$, TS$\downarrow$). This performance gap could be explained by the presence of more natural noises in VGGSound and the involvement of audios from a broader range of categories compared to speech.

The distributions of the initial and final SNR in Figure \ref{fig:snr} tell a similar story. We also calculated the ratios of mixtures enjoying an SNR improvement for all tasks. In general, both LCRs remix correctly following the text prompt. This proficiency is evident as, depending on the specific task, between 84.6\% and 99.8\% of the mixtures exhibited a positive SNRi. The ratio aligns with the average performance of the task. Interestingly, although LCR-M has a higher average performance than LCR-T in speech tasks, LCR-T remixes the speeches more accurately with higher SNRi$>$0 ratios.

The SNR of the unprocessed mixtures may influence LCR's remixing performance as well, as noisier mixtures are intuitively harder to remix. Figure \ref{fig:snr_xy} provides a micro view of all individual SNR$_{0}$ and SNRi points, underlying the macro distribution in Figure \ref{fig:snr}. Overall, SNRi does not show a statistically significant negative correlation with decreasing SNR$_{0}$. In fact, lower SNR$_{0}$ values often lead to slightly higher SNRi, as these noisier mixtures allow more room for enhancement. This highlights LCR's robustness in noisy environments,  although the final SNR (SNR$_{0}$ + SNRi) to the listener's ears remains lower for mixtures with a lower initial SNR, as expected.

\begin{table*}[!ht]
\caption{Zero-shot remixing SNRi (dB) for sound mixtures with unseen source composition, LCR-M$\vert$LCR-T. The better scores are shown in bold for each task and source composition.}.
\label{tab:zs}
\begin{center}
\begin{sc}
\begin{adjustbox}{width=\textwidth,center}
\begin{tabular}{c*{8}{c}}
\toprule
\textbf{Sound Mixtures} & \textbf{TSE} & \textbf{TSR} & \textbf{TS$\uparrow$} & \textbf{TS$\downarrow$} & \textbf{TAE} & \textbf{TAR} & \textbf{TA$\uparrow$} & \textbf{TA$\downarrow$} \\
\midrule
2 Speech & 10.1$\vert$\textbf{12.6} & n/a & 10.0$\vert$\textbf{12.7} & 9.2$\vert$\textbf{12.7} & n/a & n/a & n/a & n/a \\
2 Audio & n/a & n/a & n/a & n/a & \textbf{5.1}$\vert$4.3 & n/a & \textbf{5.3}$\vert$4.3 & \textbf{6.1}$\vert$5.3 \\
2 Speech + 1 Audio & \textbf{10.4}$\vert$9.7 & \textbf{7.8}$\vert$7.1 & \textbf{10.5}$\vert$9.6 & \textbf{10.7}$\vert$10.0 & n/a & n/a & n/a & n/a \\
1 Speech + 2 Audio & n/a & n/a & n/a & n/a & \textbf{7.9}$\vert$7.4 & \textbf{4.4}$\vert$3.6 & \textbf{5.6}$\vert$5.0 & \textbf{5.2}$\vert$4.7 \\
\hline
2 Speech + 2 Audio (FSD50K, all) & \textbf{14.0}$\vert$13.5 & 8.7$\vert$\textbf{8.9} & \textbf{9.2}$\vert$8.8 & 8.7$\vert$\textbf{8.9} & 9.0$\vert$\textbf{9.3} & 4.4$\vert$\textbf{5.0} & 4.7$\vert$\textbf{4.9} & 4.7$\vert$\textbf{5.4} \\
2 Speech + 2 Audio (FSD50K, unseen) & 13.2$\vert$\textbf{13.4} & 8.9$\vert$\textbf{9.5} & 8.7$\vert$\textbf{9.3} & 8.8$\vert$\textbf{9.2} & 8.7$\vert$\textbf{9.0} & \textbf{3.8}$\vert$3.3 & 2.8$\vert$\textbf{3.9} & 3.3$\vert$\textbf{4.4} \\
\hline
2 Speech + 2 Audio (VGGSound) & \textbf{13.7}$\vert$13.1 & \textbf{8.4}$\vert$7.9 & \textbf{8.8}$\vert$8.1 & \textbf{9.1}$\vert$8.3 & \textbf{10.3}$\vert$10.1 & 4.5$\vert$4.5 & \textbf{5.1}$\vert$4.8 & \textbf{5.0}$\vert$4.8 \\
\bottomrule
\multicolumn{9}{c}{Continued} \\
\toprule
\textbf{Sound Mixtures} & \textbf{SE} & \textbf{SR} & \textbf{S$\uparrow$} & \textbf{S$\downarrow$} & \textbf{ME} & \textbf{MVC} & \textbf{MEVC} & \textbf{OVC} \\
\midrule
2 Speech & n/a & n/a & n/a & n/a & n/a & 7.8$\vert$\textbf{10.3} & 9.8$\vert$\textbf{12.0} & 30.5$\vert$\textbf{37.8} \\
2 Audio & n/a & n/a & n/a & n/a & n/a & \textbf{4.3}$\vert$3.4 & \textbf{6.7}$\vert$5.2 & \textbf{34.8}$\vert$25.9 \\
2 Speech + 1 Audio & \textbf{9.8}$\vert$9.4 & \textbf{13.6}$\vert$12.8 & \textbf{11.8}$\vert$10.9 & \textbf{10.6}$\vert$9.8 & n/a & \textbf{7.4}$\vert$7.0 & \textbf{8.5}$\vert$8.1 & \textbf{33.9}$\vert$29.5 \\
1 Speech + 2 Audio & \textbf{14.4}$\vert$11.8 & \textbf{10.8}$\vert$8.4 & \textbf{11.9}$\vert$9.2 & \textbf{11.9}$\vert$9.7 & n/a & \textbf{5.7}$\vert$4.9 & \textbf{7.4}$\vert$6.4 & \textbf{37.1}$\vert$28.3 \\
\hline
2 Speech + 2 Audio (FSD50K, all) & 12.1$\vert$\textbf{12.4} & 12.7$\vert$\textbf{12.8} & 11.1$\vert$\textbf{11.4} & \textbf{11.1}$\vert$\textbf{11.1} & 5.2$\vert$\textbf{5.9} & 6.4$\vert$\textbf{6.8} & 7.4$\vert$\textbf{7.6} & \textbf{50.0}$\vert$47.2 \\
2 Speech + 2 Audio (FSD50K, unseen) & 12.3$\vert$\textbf{12.6} & 12.4$\vert$\textbf{12.5} & 10.6$\vert$\textbf{11.5} & 11.1$\vert$\textbf{11.7} & 4.9$\vert$\textbf{5.8} & 5.3$\vert$\textbf{6.0} & 6.4$\vert$\textbf{7.2} & 47.6$\vert$47.6 \\
\hline
2 Speech + 2 Audio (VGGSound) & \textbf{11.7}$\vert$10.7 & \textbf{12.0}$\vert$11.1 & \textbf{11.1}$\vert$10.2 & \textbf{10.6}$\vert$9.6 & 6.5$\vert$6.5 & \textbf{6.6}$\vert$6.3 & \textbf{7.2}$\vert$6.8 & \textbf{52.1}$\vert$44.2 \\
\bottomrule
\end{tabular}
\end{adjustbox}
\end{sc}
\end{center}
\end{table*}

\subsection{Zero-shot Evaluation} \label{sec:zse}

Table \ref{tab:zs} presents the zero-shot performance of LCR-M and LCR-T on sound mixtures with different number of sources (rows 1 to 4)\footnote{Some entries are filled with N/A because some tasks are not defined (e.g. TAE for 2 Speech) or are equivalent to another task (e.g. TSE = TSR, both resulting in one speech left, for 2 Speech).} or including unseen audio sources (rows 5, 6), compared to the in-domain performance (the last row). The performance on 2 Speech and 2 Speech + 1 Audio mixtures are comparable to the performance of 2 Speech + 2 Audio mixtures used for training, while the performance on 1 Speech + 2 Audio and 2 Audio mixtures are worse but still at least 3.4 dB better than the original mixtures. The ability to generalize to unseen numbers of sources is partly due to the audio sources from VGGSound not being perfectly clean and of the same length. For instance, there may be multiple sources with the same label, unlabeled background noise, or occasional silent periods, effectively resulting in more or fewer active sources during training. For similar reasons, the zero-shot performance on audio sources from FSD50K is even better in most tasks compared to VGGSound, possibly because FSD50K has higher audio quality and some audio labels are covered in VGGSound. To eliminate the latter factor, we evaluated LCR on audios with strictly unseen labels\footnote{Synonyms were considered for excluding seen labels.} from FSD50K and observed a 0.3 to 1.9 dB performance drop on audio tasks (TAE, TAR, TA$\uparrow$, TA$\downarrow$) but similar performance on the rest. Interestingly, LCR-T shows a better zero-shot performance than LCR-M in 2 Speech and unseen audio remixing than LCR-M, although the latter has a higher in-domain performance. Therefore, we evaluated (and finetuned) LCR-T for the experiments in the remaining section, although both models have shown remarkable zero-shot remixing performance on mixtures with unseen compositions.

\begin{table}[!t]
\centering
\caption{Zero-shot speech remixing SNRi (dB) for two speakers with one or multiple style variation.}
\label{tab:2spk_style}
\begin{sc}
\begin{adjustbox}{width=0.8\linewidth,center}
\begin{tabular}{c*{5}{c}}
\toprule
Differ in & TSE & TS$\uparrow$ & TS$\downarrow$ & MVC & Average \\
\midrule
Gender & 11.4 & 12.5 & 13.0 & 10.3 & 11.8 \\
Pitch & 11.3 & 11.7 & 11.6 & 8.7 & 10.8 \\
Tempo & 6.7 & 5.4 & 6.6 & 3.3 & 5.5 \\
Energy & 10.9 & 11.5 & 11.8 & 10.7 & 11.2 \\
Emotion & 11.4 & 11.9 & 11.6 & 9.6 & 11.1  \\ 
\midrule
Multiple & 13.2 & 13.2 & 13.0 & 10.6 & 12.5 \\
\bottomrule
\end{tabular}
\end{adjustbox}
\end{sc}
\end{table}

\begin{table}[!t]
\caption{Zero-shot and finetuned TSE SI-SDR (dB) for two speakers differing in gender or energy, compared to LLM-TSE (scores in parentheses obtained with an additional audio clue).}
\label{tab:tencent}
\centering
\begin{sc}
\begin{adjustbox}{width=\linewidth,center}
\begin{tabular}{cccc}
\toprule
Differ in & zero-shot & 2.8-hour Finetuned & LLM-TSE \cite{hao2023typing} \\
\midrule
Gender & 8.4 & 12.1 & 10.4 (10.9) \\
Energy & 7.0 & 10.8 & 8.9 (9.4) \\
\bottomrule
\end{tabular}
\end{adjustbox}
\end{sc}
\end{table}

\setlength{\tabcolsep}{1pt}
\begin{table}[!t]
\centering
\caption{Zero-shot subjective evaluation on AudioSet samples. F: prompt following Q: audio quality}
\label{tab:audioset_zs}
\normalsize
\begin{sc}
\begin{adjustbox}{width=\linewidth,center}
\begin{tabular}{c|cc|cc|cc|cc}
\toprule
\multirow{2}{*}{\diagbox[dir=NW]{Model}{Task}} & \multicolumn{2}{c|}{\textbf{Enhance}}  & \multicolumn{2}{c|}{\textbf{Extract}} & \multicolumn{2}{c|}{\textbf{Remove}} & \multicolumn{2}{c}{\textbf{Multiple}} \\
 & MOS-F & MOS-Q & MOS-F & MOS-Q & MOS-F & MOS-Q & MOS-F & MOS-Q \\
\midrule
Sepformer \cite{sepformer} & 2.71 & 3.01 & - & -  & - & - & - & - \\
AudioSep \cite{liu2023separate} & 2.95 & 3.15 & 2.82 & 3.23  & 3.20 & 3.57 & 2.15 & 3.34 \\ \hline
LCR-T  & \bf{3.77} & \bf{3.63} & \bf{3.59} & \bf{3.48}  & \bf{3.53} & \bf{3.67} & \bf{2.88} & \bf{3.42} \\
\bottomrule
\end{tabular}
\end{adjustbox}
\end{sc}
\end{table}

Table \ref{tab:2spk_style} further measures the zero-shot two-speaker remixing performance of two speakers with controlled style difference to analyze the influence of each style attribute. Our results show that LCR-T can effectively remix speakers as long as they have one distinct style attribute. A difference in gender, energy, pitch, or emotion results in an average 5.3 dB or higher SNRi compared to a difference in tempo, which requires a longer window to calculate the speaking speed. LCR's ability to remix the target speaker(s) is further improved when two speakers differ in multiple style attributes. Table \ref{tab:tencent} compares our performance with the target speech extraction performance reported in LLM-TSE \cite{hao2023typing} in gender and energy evaluated on similar two-speaker mixtures\footnote{Mixtures were generated with the same SNR distribution.}. The performance is measured in Scale-Invariant Signal-to-Distortion ratio (SI-SDR) \cite{sisdr}. While LCR's zero-shot performance falls behind that of LLM-TSE by 2 dB, after finetuning with 2k two-speaker mixtures (equivalent to 2.8 hours), LCR surpasses LLM-TSE, even if an additional audio clue is provided to the latter. This demonstrates that LCR, as a versatile soundscape remixer, can outperform an expert model in a specific task after finetuning with limited data. 

Finally, we performed subjective evaluation using real recordings from AudioSet \cite{audioset} (unseen neither as a mixture nor as a mixing source). For each sample, we wrote four text prompts corresponding to speech enhancement (SE), target audio/speech extraction (TAE or TSE), target audio/speech removal (TAR or TSR), and multiple sound extraction or removal (ME). We then asked participants to rate how well the remixed mixture followed (F) the text prompt and the quality (Q) of the remixed mixture. We report mean opinion scores MOS-F and MOS-Q in Table \ref{tab:audioset_zs}. More details can be found in Appendix \ref{app:human}. Two baseline models are compared: a Sepformer \cite{sepformer} model trained for noise suppression \cite{dubey2022icassp}\footnote{Sepformer: huggingface.co/speechbrain/sepformer-dns4-16k-enhancement} and the AudioSep \cite{liu2023separate}\footnote{AudioSep: https://github.com/Audio-AGI/AudioSep} model trained for target audio extraction. While AudioSep can only handle speech enhancement (with `speech' as the prompt) and extraction tasks, we subtracted the extraction result from the mixture for the removal task and ran AudioSep multiple times for the multiple task. In contrast, LCR always remixed in a single run. Table \ref{tab:audioset_zs} shows LCR-T significantly wins both the Sepformer denoiser and AudioSep by a large margin in both prompt following and audio quality. This can be attributed to the diversity of our curated training data, which better matches the in-the-wild acoustic distribution, and our use of rephrased text prompts that consider different expressions of the same sounds and instructions. In contrast, the DNS4 dataset \cite{DNS4} used to train Sepformer contains fewer mixtures with high-energy audio sources. Additionally, LCR-T's transformer-based architecture may offer an advantage over AudioSep's CNN-based architecture. Our results demonstrate that LCR trained on synthetic mixtures can generalize to in-the-wild sound mixtures.

\subsection{Visualization of Semantic and Acoustic Filters} \label{sec:analysis}
\begin{figure*}[ht]
\centering
\includegraphics[width=\textwidth]{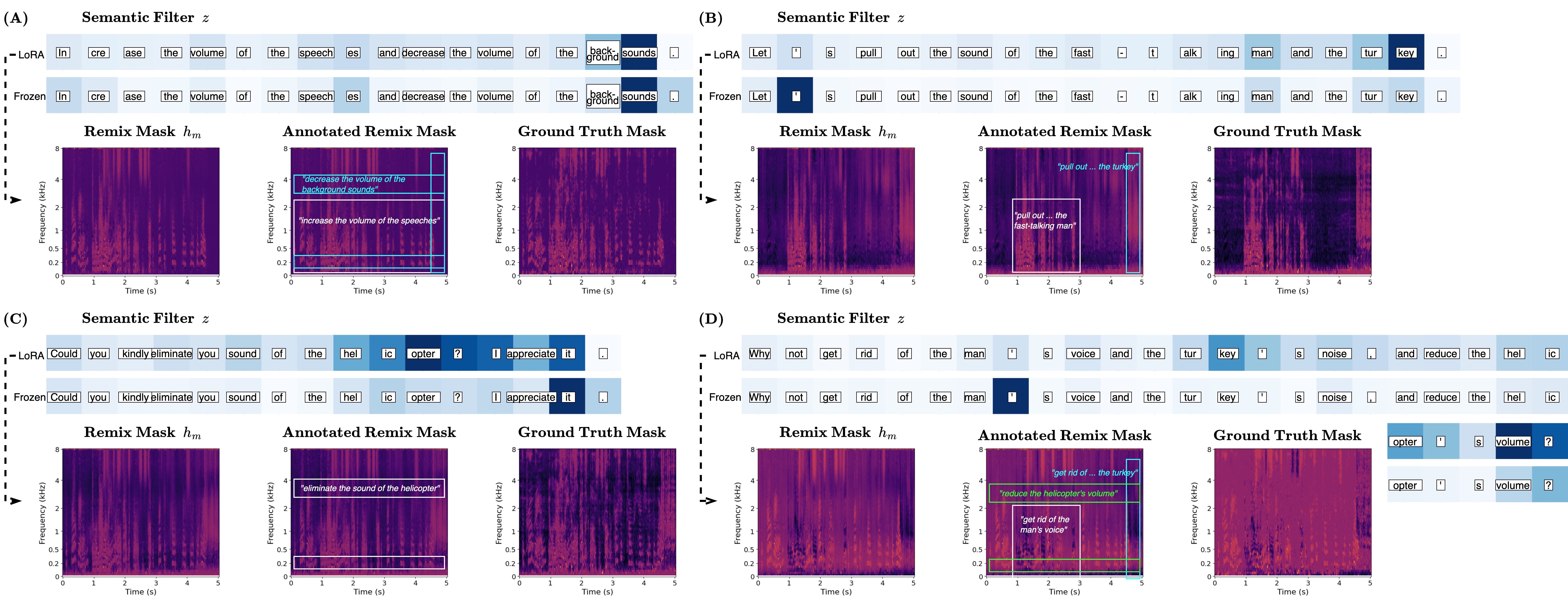}
\caption{The attention scores of semantic filter $z$ when reading text prompts and the resulting remixing mask $h_m$. The attention scores are calculated at the last \texttt{EOS} token, ignoring token \texttt{SOS} and \texttt{EOS} itself, and darker colors indicate larger scores.}
\label{fig:features}
\end{figure*}

To better understand the behavior of LCR(-T), we visualized the semantic filter $z$ calculated by the \textit{PromptReader} when reading the text prompt and the remixing mask (acoustic filter) $h_m$ calculated by the \textit{SoundRemixer} from $z$. The attention scores of the last layer of \textit{PromptReader} LLaMA2 (frozen or LoRA finetuned) when reading text prompts, and the remixing mask of the \textit{SoundRemixer} compared to the ground truth one are plotted in Figure \ref{fig:features}. Readers can refer to Figure \ref{fig:spec1} for the remixed spectrograms produced by these filters.

Attention visualization reveals that \textit{acoustic keywords} are highlighted in each text prompt. These keywords include ``background sounds",  ``helicopter", ``man", ``turkey", ``volume", etc. We also observe that the LoRA-finetuned \textit{PromptReader} pays more attention to these keywords than a non-finetuned (frozen) one. Greater attention suggests that the semantic filter generated by a finetuned \textit{PromptReader} better encodes the sound objects mentioned in the text prompt, guiding the downstream \textit{SoundRemixer} to estimate a more accurate remixing mask, which we also annotate the amplification or suppression patterns of low-frequency man's voice and high-frequency helicopter's noise. This finding provides an insight into our experimental result that a finetuned \textit{PromptReader} outperforms a frozen one by more than 4.5 dB in Table \ref{tab:all}.

\begin{table}[!t]
\caption{Ablations on the text prompt quality, the training method, and the model configuration with LCR-C.}
\label{tab:abl}
\begin{center}
\begin{sc}
\begin{adjustbox}{width=\linewidth,center}
\begin{tabular}{cccccccccc}
\toprule
Setting & TSE & TSR & TS$\uparrow$ & TS$\downarrow$ & TAE & TAR & TA$\uparrow$ & TA$\downarrow$ & SE \\
\midrule
Fixed Prompts & 7.7 & 3.4 & 4.2 & 2.2 & 6.4 & 2.1 & 2.5 & 1.6 & 7.5 \\
MultiMask & 9.6 & 4.4 & 4.6 & 4.7 & 8.3 & 2.5 & 3.1 & 3.0 & 8.2 \\
MM + PIT & 9.7 & 4.6 & 4.8 & 4.9 & 8.2 & 2.5 & 3.1 & 3.0 & 8.3 \\
\hline
Kernel $4\times$  & 9.5 & 4.3 & 4.7 & 4.7 & 8.3 & 2.7 & 3.3 & 3.3 & 8.1\\
Causal & 8.5 & 3.3 & 3.3 & 3.4 & 7.7 & 1.9 & 2.5 & 2.5 & 6.9 \\
\hline
Default & 9.7 & 4.6 & 4.7 & 4.8 & 8.2 & 2.5 & 3.1 & 3.0 & 8.2 \\
\bottomrule
\multicolumn{10}{c}{Continued} \\
\toprule
SETTING & SR & S$\uparrow$ & S$\downarrow$ & ME & MVC & MEVC & OVC & \multicolumn{2}{c}{Average} \\
\midrule
Fixed Prompts & 7.7 & 6.9 & 6.4 & 3.6 & 3.2 & 4.0 & 32.6 & \multicolumn{2}{c}{6.4} \\
\hline
MultiMask & 8.5 & 7.4 & 7.0 & 4.3 & 3.8 & 4.6 & 42.5 &  \multicolumn{2}{c}{7.9} \\
MM + PIT & 8.6 & 7.2 & 6.9 & 4.3 & 4.0 & 4.7 & 42.6 & \multicolumn{2}{c}{8.0} \\
\hline
Kernel $4\times$ & 8.5 & 7.5 & 7.1 & 4.3 & 3.9 & 4.5 & 36.1 & \multicolumn{2}{c}{7.6} \\
Causal & 7.2 & 6.0 & 5.7 & 3.7 & 3.2 & 3.8 & 37.2 & \multicolumn{2}{c}{6.7} \\
\hline
Default & 8.5 & 7.4 & 7.0 & 4.2 & 3.9 & 4.6 & 40.3 & \multicolumn{2}{c}{7.8} \\

\bottomrule
\end{tabular}
\end{adjustbox}
\end{sc}
\end{center}
\end{table}

\subsection{Ablations} \label{sec:ablations}
We conducted all ablations using the LCR-C (with finetuned LLaMA2) model due to computational constraints as documented in Table \ref{tab:abl}. The first row shows that training on fixed prompts (without rephrasing by ChatGPT) significantly degrades the performance by around 1.4 dB on average. The second to the last row studies different model configurations: estimating multiple masks for all sources, multiple masks with additional PIT loss, a kernel and stride size 4 times larger, and a causal implementation with causal TCNs. Notice that estimating multiple masks does not improve the performance, so one remixing mask is sufficient. The causal model runs around 90 times faster than real-time on an NVIDIA L40 GPU with pre-computed semantic filters.

\section{Conclusion and Limitation}

We develop \textit{Listen, Chat, and Remix}, the first text-guided sound enhancement system that can arbitrarily remix every source in speech and audio mixtures. Our development also introduces the first text-prompted sound remixing dataset of 160 hours, featuring speakers with diverse styles, audio from hundreds of classes, and five generic text prompts written for each mixture. LCR trained on this dataset demonstrates an SNR improvement of over 11 dB averaged across all 16 tasks, superior performance than ad-hoc sound extraction models, and a generalization ability to unfamiliar mixtures of unseen sound types and source numbers. Although LCR can remix in-the-wild sound mixtures, it has yet to be generalizable to all real-world scenarios. Challenges include very low signal-to-noise ratios, a large number of sound sources, and new types of sounds that are acoustically different from any sounds LCR has trained on. While our dataset tries to capture diverse sources and mixing conditions, real-world acoustic environments exhibit unpredictable variations that are difficult to replicate synthetically. Live recordings with natural soundscapes would be a favorable direction to further improve robustness and realism. Therefore, future efforts will focus on scaling up the model size and dataset to achieve better performance and better generalization to more challenging cases.

\section*{Acknowledgments}
This work was funded by the National Institutes of Health
(NIH-NIDCD) and a grant from Marie-Josee and Henry
R. Kravis. We would like to thank Gavin Mischler for
suggesting the name Listen, Chat, and Remix.

\bibliographystyle{IEEEtran}
\bibliography{refs}

\vspace{-0.2cm}

{\appendices

\section{Models and Training} \label{app:training}

The remix block of \textit{SoundRemixer} can be either a TCN, transformer, or Mamba. We adopted the implementation of SpeechBrain \cite{SpeechBrain} for the former two and the official implementation of Mamba-TasNet (L)\footnote{Mamba-TasNet (L): https://github.com/xi-j/Mamba-TasNet} the third. We followed the default configurations) as their papers \cite{convtasnet, sepformer, mambatasnet}. The \textit{SoundRemixer} was trained from scratch, while the \textit{PromptReader} was finetuned from pretrained checkpoints\footnote{LLaMA2: https://huggingface.co/meta-llama/Llama-2-7b-chat-hf. GPT-2: https://huggingface.co/gpt2}. We trained all LCRs for 100 epochs using 4 NVIDIA L40 GPUs with bf1 precision. The total batch size was set to 16 for LCR-C or 8 for LCR-T and LCR-M. We used an Adam \cite{adam} optimizer with a learning rate of $5e-4$ for TCN (LCR-C) or $1e-4$ for transformer (LCR-T) and Mamba (LCR-M). GPT-2 and LLaMA2 were finetuned with a learning rate of $1e-4$. GPT-2 was finetuned with all parameters, and LLaMA2 was finetuned using LoRA \cite{hu2022lora} with a rank of 16 and a dropout rate of 0.05 for the query and value matrices in each self-attention layer. In addition, we applied a linear learning rate warm-up for the first 5000 updates, and in case the validation Signal-to-Noise Ratio (SNR) did not improve for 3 epochs, we halved the learning rate for both the \textit{SoundRemixer} and \textit{PromptReader}.

\vspace{-0.5cm}

\section{Human evaluation} \label{app:human}
To ensure the quality and relevance of our evaluations, we required raters to be native English speakers living in the United States. We applied the following filters:
\begin{itemize}
    \item HIT Approval Rate (\%) for all Requesters' HITS: \verb|greater than 95|.
    \item Location: \verb|is UNITED STATES (US)|.
    \item Number of HITs Approved: \verb|greater than 50|.
\end{itemize}
We conducted four batches of surveys to evaluate different aspects of our model's performance: Speech Enhancement, Target Speech/Audio Extraction, Target Speech/Audio Removal, and Multiple Extraction or Removal. Each survey contained 20 sets of audio to be rated, and we collected ratings from 10 subjects. Subjects were asked to rate whether the remixed audio adhered to the text instructions and how good the sound quality was, both on a scale of 5. We randomly permuted the order of the remixed audio samples in each set without revealing any information to the subjects. On average, each subject completed the survey in 18 minutes, and we compensated them with 10 dollars each.


}





\vfill

\end{document}